\documentstyle[10pt,aaspp4,epsfig]{article}

\newcommand{\kpoh}{{k+\frac{1}{2}}}
\newcommand{\npo}{{n+1}}
\newcommand\simgreater{\,\lower0.7ex\hbox{$\stackrel{>}{\sim}$}\,}
\newcommand\simless{\,\lower0.7ex\hbox{$\stackrel{<}{\sim}$}\,}
\newcommand\bldomega{\mbox{\boldmath $\omega$}}

\slugcomment{Submitted to Ap. J.}

\lefthead{Swesty,Wang, and Calder}

\begin{document}

\title{Numerical Models of Binary Neutron Star System Mergers. I.:
Numerical Methods and Equilibrium Data for Newtonian Models}

\author{F. Douglas Swesty\altaffilmark{1,2,3}, Edward Y. M. Wang
\altaffilmark{1,3},
and Alan C. Calder\altaffilmark{3,4}}

\altaffiltext{1}{Department of Physics and Astronomy,
SUNY at Stony Brook, NY 11794}

\altaffiltext{2}{Department of Astronomy,
University  of Illinois, Urbana, IL 61801}

\altaffiltext{3}{National Center for Supercomputing
Applications, University  of Illinois, Urbana, IL 61801}

\altaffiltext{4}{ASCI/Alliances Center for Astrophysical 
Thermonuclear Flashes, Department of Astronomy and 
Astrophysics,
University of Chicago, Chicago, IL 60637}

\begin{abstract}
The numerical modeling of binary neutron star mergers has become a
subject of much interest in recent years.  While a full and accurate
model of this phenomenon would require the evolution of the equations
of relativistic hydrodynamics along with the Einstein field equations,
a qualitative study of the early stages on inspiral can be
accomplished by either Newtonian or post-Newtonian models, which are
more tractable.  However, even purely Newtonian models present
numerical challenges that must be overcome in order to have accurate
models of the inspiral.  In particular, the simulations must maintain
conservation of both energy and momenta, and otherwise exhibit good
numerical behavior.  A spate of recent papers have detailed the
results for Newtonian and post-Newtonian models of neutron star
coalescence from a variety of groups who employ very different
numerical schemes.  These include calculations that have been carried
out in both inertial and rotating frames, as well as calculations that
employ both equilibrium configurations and spherical stars as initial
data.  However, scant attention has been given to the issue of the the
accuracy of the models and the dependence of the results on the
computational frame and the initial data.  In this paper we offer a
comparison of results from both rotating and non-rotating (inertial)
frame calculations.  We find that the rotating frame calculations
offer significantly improved accuracy as compared with the inertial
frame models.  Furthermore, we show that inertial frame models exhibit
significant and erroneous angular momentum loss during the simulations
that leads to an unphysical inspiral of the two neutron stars.  We
also examine the dependence of the models on initial conditions by
considering initial configurations that consist of spherical neutron
stars as well as stars that are in equilibrium and which are tidally
distorted.  We compare our models those of Rasio \& Shapiro
(1992,1994a)\nocite{rassha92,rassha94} and New \& Tohline
(1997)\nocite{nt97}.  Finally, we investigate the use of the isolated
star approximation for the construction of initial data.

\end{abstract}


\pagebreak

\section{Introduction}

\subsection{Scientific motivation}

Binary Neutron Star Mergers (NSMs) are unique laboratories for the
study of astrophysics.  The merger involves many elements of the
theory of relativistic astrophysics, gravitational wave astronomy, and
nuclear astrophysics. Furthermore, NSMs are thought to be a possible
source of detectable gravitational radiation, the {\it r}-process
elements, and gamma ray bursts.  In order to develop accurate
models of NSMs one must numerically solve the equations describing gas
dynamics and the gravitational field arising from the matter.
However, the need for accurate numerical models of the inspiraling
binary system presents some unique challenges that we address in this
paper.

In realistic astrophysical situations the merger of binary neutron star
systems is driven by gravitational radiation losses (Misner et al. 1973) 
\nocite{mtw73}.  This loss of energy will lead to the inspiral and
eventual coalescence of the binary system.  The prediction of energy
loss by gravitational radiation was confirmed by the observation of
PSR1913+16, a binary neutron star system
(Hulse \& Taylor 1975)\nocite{hulse75}.  The observed rate of
decrease of the orbital period of this system is in good agreement
with predictions made by general relativity (Taylor \& Weisberg
1989)\nocite{taylorweisberg89}.  Coalescing binary systems are
expected to emit tremendous amounts of energy in the form of
gravitational waves during the final stages of coalescence, and the
gravitational waves produced in these events are expected to be
observed by gravitational wave detectors currently under construction.
Gravitational wave interferometers such as LIGO (Abramovici et al.
1992)\nocite{abramovici92} and VIRGO (Bradaschia et al.
1990)\nocite{bradaschia90} will soon be operational and will present
the first opportunities to study NSMs via gravitational waves.
Theoretical templates of the expected signal are required to extract
signal information from the noisy background (Cutler et al.
1993)\nocite{cutler93}.  Post-Newtonian methods (Lincoln \& Will
1990)\nocite{lincoln90} may be adequate for the prediction of
waveforms for the early stage of the inspiral. However, the
prediction of waveforms in the later stages of the merger, when tidal
effects and neutron star structure become important, requires a full
three-dimensional numerical solution of the equations describing the
motion of matter and the gravitational field.

In addition to gravitational wave astronomy, NSMs are of interest for
nuclear astrophysical reasons.  NSMs may yield information about the
structure of neutron stars.  Since the equation of state (EOS) of
neutron star matter is not well constrained, the observation of
gravitational wave signals from NSMs may provide constraints that
could provide information about the dynamics of the merger and in turn
the EOS of dense matter.  Additionally, the material ejected during
the coalescence of binary neutron stars may be a site of {\it
r}-process nucleosynthesis (Lattimer et al. 1977)\nocite{lmrs77}. The
{\it r}-process, which is thought to be responsible for the production
of about 50\% of elements heavier than iron in the universe, occurs
when the capture rate of neutrons bombarding nuclei exceeds the beta
decay rate.  In the material ejected during NSMs there are expected to
be regions where 
the {\it r}-process occurs robustly (Meyer 1989)\nocite{meyer89}.
Simulations of NSMs can allow us to study both the mass ejection 
and nucleosynthesis that occurs in the ejected material.

NSMs are a suggested source for the mysterious gamma-ray bursts
observed by CGRO and other high-energy observational missions.  NSMs
are thought to release energy on the order of their gravitational
binding energy $\approx 10^{53}$ erg, which may be larger than
estimated gamma-ray burst energies $\approx 10^{51}$ (Quashnock
1996,Rees 1997)\nocite{quashnock96,rees97a} to $10^{53}$ erg (Woods \&
Loeb 1994)\nocite{woods94}.  A popular model for bursts at
cosmological distances is the relativistic fire-ball (Paczy\'{n}ski
1986, Goodman 1986, Shemi \& Piran 1990, Paczy\'{n}ski
1990)\nocite{paczyn86,goodman86,shemi90,paczyn90}.  NSMs are likely
candidates for the source of relativistic fire-balls, but the
mechanism by which the fire-ball develops has yet to be
determined. Observations in the spring of 1997 of optical and X-ray
counterparts to GRB 970228 (Costa, et al. 1997, Guarnierni 1997,
Piro et al. 1997)\nocite{costa97a,guarnieri97a,piro97a} and GRB 970508
(Bond 1997, Djorgovski et al. 1997, Metzger et al.
1997)\nocite{bond97a,djorgovski97a,metzger97a}, particularly the
measurement of a Mg II absorption line at redshift $z = 0.835$
(Metzger et al. 1997), suggest that the bursts do indeed have a
cosmological origin. Simulations of NSMs can test the consistency of
the energetics and time scales with the estimated energies and
observed time scales of the observed bursts.

One of the major difficulties in carrying out numerical simulations of
binary neutron star mergers is developing a numerical algorithm that
does not introduce unphysical dynamical effects into the problem.  In
order to avoid spurious inspiral, both the total energy and angular
momentum must be conserved to sufficient accuracy.  In the absence of
physical instabilities or dissipative effects, such as gravitational
radiation losses, the numerical methods should be capable of
maintaining a binary neutron star system in a stable orbit.
Additionally, when physical instabilities capable of causing
coalescence are present, the algorithms must continue to conserve all
important physical quantities.  Without this capability it is
impossible to develop quantitatively accurate models in situations
where radiation losses are present.  In this paper we consider several
variations on the popular ZEUS hydrodynamic algorithm (Stone \&
Norman 1992) as applied to NSMs. The comparisons that will appear
later in this paper examine the numerical effects of the choice of
rotating versus inertial frames as well as the choice of several
possible schemes for the coupling of gravity to the hydrodynamics.
This paper is intended to lay the numerical groundwork for
post-Newtonian and relativistic studies that will follow in later
papers.  While realistic models of NSMs are clearly relativistic
or, at a bare minimum, post-Newtonian (PN) in nature, the examination
of Newtonian models of orbiting binaries neutron stars is still
of considerable value.  Many of the lessons learned from Newtonian
models will provide guidance for PN or GR modeling efforts.
Indeed, if a numerical self-gravitating hydrodynamics algorithm is 
incapable of maintaining stable orbits for binary star systems in the 
Newtonian limit, then it is unlikely to be useful for more complex,
and realistic, simulations of NSMs.   In this paper we concentrate 
on purely Newtonian models of orbiting binary neutron stars in both
the stable and unstable regimes.  We will consider the evolution of
initial  configurations that are tidally unstable as well as 
initial configurations involving both spherical and ``relaxed'' 
neutron stars.  Post-Newtonian models that make use of our numerical 
techniques will be considered in a subsequent paper.

\subsection{Status of contemporary work on Newtonian and
PN simulations of binary neutron star systems}

The Newtonian and PN simulations that have been carried out to date
can be placed into two categories: those that have employed Eulerian
hydrodynamic methods (see Bowers \& Wilson 1991 for a discussion of
Eulerian methods)\nocite{bw91} and those that have employed smoothed
particle hydrodynamics (SPH) methods (see Gingold \& Monaghan 1977 or
Hernquist \& Katz 1989 for a discussion of SPH
methods)\nocite{lucy77,gingold77,hernquistkatz89}, an inherently
Lagrangean technique.  Additionally, these simulations have been
carried out in both inertial, i.e. laboratory, frames and in
non-inertial, rotating frames.  These simulations have also utilized a
range of techniques for calculating the gravitational potential.
Finally, these simulations have employed both spherical stars and
equilibrium binary configurations as initial data.  These choices can
play a critical role in determining the outcome of the simulations.
For this reason, in this section we briefly describe existing work on
Newtonian and PN binary neutron star systems with a focus on the
numerical techniques and the initial configurations that have been
used.  We first consider the Eulerian calculations followed by the SPH
models.

The earliest Eulerian models of binary neutron star systems were
carried out by Oohara and Nakamura (1989)\nocite{oohara89}.  
This work and subsequent papers 
(Nakamura \& Oohara 1989, Oohara \&
Nakamura 1990, Nakamura \& Oohara 1990, Nakamura \& Oohara 1991,
Oohara \& Nakamura 1992)
\nocite{oohara89,nakamura89,oohara90,nakamura91}
made use of purely Newtonian hydrodynamics, while later work with
Shibata included PN effects (Oohara \& Nakamura 1992, Shibata,
Nakamura, \& Oohara 1992,1993)\nocite{oohara92,sno92,sno93,son97}.  As
with all Eulerian, i.e. grid-based, hydrodynamics methods the
underlying PDEs are discretized onto a coordinate mesh. The evolution
of the mass distribution occurs as the material flows through the grid
zones, and the equations governing the evolution are
finite-differenced analogs of the Euler equations. There are many
approaches to finite-differencing the Euler equations, but most modern
formulations are at least second order in time and space, and have
methods of realistically modeling shocks.  The hydrodynamics method
employed by the Oohara/Nakamura/Shibata calculations utilizes
LeBlanc's method for transport, making use of a tensor artificial
viscosity. A brief description is given in an appendix to Oohara and
Nakamura (1989).  The earlier calculations were carried out in the
laboratory (fixed) frame, while later models utilized a rotating
frame.  In all of the calculations the gravitational potential was
found by a direct solution of the Poisson equation.  However, none of
the papers discuss the boundary conditions for this equation.
Finally, the papers have considered both spherical stars and
equilibrium configurations for initial data although no comparisons of
the two types of initial data were offered.

Ruffert et al. (1996,1997,1997)\nocite{rjs96,rjts97,rrj96} performed
PN simulations of NSMs with the PROMETHEUS code implementing the
Piecewise Parabolic Method (PPM) of Colella and Woodward
(1984)\nocite{colw84}.  The PPM is an extension of Godunov's method
that solves the Riemann problem locally for the flow between zone
interfaces, and, accordingly, it is well suited to addressing shocks.
The calculation of the gravitational potential was accomplished by
means of the direct solution of Poisson's equation using
zero-padding boundary conditions (Hockney 1988)\nocite{hockney88}.
The initial conditions of the simulations were spherical neutron stars
in both tidally locked and rotating configurations.  The stars were
embedded in an atmosphere of $10^9$g/cm$^3$ that covered the entire
grid, and an artificial smoothing was performed on the surfaces of the
stars to soften the edges. The earlier models considered
configurations with the realistic equation of state of Lattimer and
Swesty (1991)\nocite{ls91} while later studies considered models with
a much simpler polytropic EOS (Ruffert, Rampp, \& Janka 1997).

The most recent Eulerian simulations of binary neutron star systems
were performed by New and Tohline (1997)\nocite{nt97}. Their work
focused on the evolution of equilibrium sequences of co-rotating,
equal mass pairs of polytropes. The equilibrium sequences were
constructed with Hachisu's Self-Consistent Field (SCF) technique
(Hachisu 1986a,b)\nocite{hachisu86a,hachisu86b}. The dynamical
stability of these equilibrium sequences was tested by evolving them
with a 2nd order accurate finite-differenced Newtonian hydrodynamics
code.  The gravitational potential was obtained by a direct solution
of Poisson's equation accomplished by means of the alternating
direction implicit method.  No description was given of the boundary
conditions that were applied to the Poisson equation.  The calculation
was carried out in a rotating frame that was initially co-rotating
with the binary system in order to avoid problems with the advection
of the stars across the grid.  In a stability test, a comparison of
two white dwarf binary system simulations starting from the same
initial conditions, one carried out in the inertial reference frame
and the other carried out in the initially co-rotating frame, revealed
dramatic differences in the dynamics of the binary system.  This
difference illustrates the need for very careful studies of purely
numerical effects on these types of simulations. In the stability
tests, New and Tohline found no points of instability for polytropic
models with fairly soft equations of state ($\gamma = 2, 5/3$).  They
did find, however, an instability for the stiffer $\gamma = 3$ case
indicating that systems with stiffer equations of state are
susceptible to tidal instabilities. It is worthwhile to note that the
authors state that they may have misidentified some stable systems as
unstable had they performed their simulations in the inertial
reference frame, and they therefore stress the importance of very
careful studies of numerical effects and careful comparison of
different numerical methods.

The earliest SPH simulations of binary neutron star systems began
with the work of Rasio and Shapiro 
(1992,1994,1995)\nocite{rassha92,rassha94,rassha95}.
In SPH, a distribution of mass in a particular region of space is
represented by discreet particles such that the mass density of the
particles is proportional to the specified density of the fluid. Local
calculations of fluid quantities are accomplished by smoothing over
the local distribution of particles.  The method is inherently
Lagrangean.  This numerical work accompanied semi-analytical work with
Lai, with the goal of predicting the onset of instabilities in binary
systems 
(Lai, Rasio, \& Shapiro 1993a,1993b,1993c,1994a,1994b,
1994c)\nocite{lrs93a,lrs93b,lrs93c,lrs94a,lrs94b,lrs94c}.  
The calculation of the gravitational potential was accomplished by
mapping the particle distribution onto a density distribution on a
mesh where the Poisson equation was solved by means of fast Fourier
transform methods.  In these calculations, the authors employed
several different polytropic equations of state.  The equilibrium
initial data for the models were obtained by allowing the spherical
stars to relax to a steady state solution in a rotating frame of
reference.  The work investigated sequences of binary star systems
with a range of initial separations, and their construction of
equilibrium initial configurations for evolutions was critical in
determining if and when a dynamical instability forced the merger. The
group reported the presence of a dynamical instability at separations
that increase with an increase of the polytropic exponent $\gamma$.

The Drexel group performed Newtonian SPH simulations of
nonaxisymmetric collisions of equal mass neutron stars (Centrella \&
McMillan 1993)\nocite{cm93}.  This effort was followed by calculations
of Newtonian NSMs by Zhuge et al. (1994,1996)\nocite{zcm94,zcm96} They
employed the TREESPH implementation of SPH developed by Hernquist and
Katz (1989) in which the gravitational forces are calculated via the
hierarchical tree method of Barnes and Hut (1986)\nocite{barneshut86}.
The calculations employed a polytropic EOS with $\gamma = 5/3$ and
$\gamma = 2$.  These calculations were carried out in the laboratory
frame and the initial data consisted of spherical stars in the
non-rotating case or rotating stars that were produced using a
self-consistent field method.  The work was aimed at studying the
gravitational radiation emission from NSMs and addressed effects due
to the equation of state, spins, and mass ratio of the stars on the
gravitational wave energy spectrum.  The coalescence in these
calculations was driven by a frictional force term added to the
hydrodynamic equations that models the effects of gravitational
radiation loss.
 
Davies, Benz, Piran, and Thielemann (1994)\nocite{dbpt94} performed
SPH simulations of NSMs with a focus on the nuclear astrophysical and
thermodynamic effects of coalescence. The SPH code used for these
calculations was described in earlier work on stellar collisions (Benz
\& Hills 1987)\nocite{benzhills87}, and it makes use of a tree
algorithm for calculating gravitational forces.  The calculations were
carried out in the inertial frame.  The initial data for the neutron
stars was modeled as equal mass $\gamma = 2.4$ polytropes, but a more
realistic EOS was employed for the dynamical calculation.  The driving
force behind the coalescence was a frictional force model of
gravitational radiation loss similar to that of the Drexel group.  The
rates of energy and angular momentum loss were determined by applying
the quadrupole approximation to the equivalent point mass system, and
the resulting acceleration of each SPH particle was determined by
expressions derived from these rates.  Subsequent calculations by
Rosswog et al. (1998)\nocite{rosswog98} have focused on {\it
r}-process nucleosynthesis and mass ejection.  More recently members
of this group have developed a PN extension of the SPH algorithm (Ayal
et al. 1999)\nocite{ayal99}, which they have applied to NSMs to study
the dynamics and gravitational wave emission from the merger.

\subsection{Outline}

In the remainder of this paper we will focus on comparisons of several
Eulerian numerical methods for modeling Newtonian binary neutron star
systems as well as comparisons of the effects of spherical versus
equilibrium initial data.  Additionally, we will examine the stability
of both the spherical and equilibrium initial data.
The subject of gravitational wave signals from NSMs will not be 
considered here, but instead will be the subject of a subsequent paper.

In \S\ref{sec:hydro} we explain our numerical methods for evolving the
equations of hydrodynamics and solving the Poisson equation.  In
\S\ref{sec:eqd} we delineate our method for obtaining equilibrium data
with self-consistent boundary conditions.  In \S\ref{sec:self} we
compare calculations carried out in both rotating and inertial frames
with several different schemes for coupling gravity to matter via the
gas momentum equations.  In \S\ref{sec:stab} we compare models with
both equilibrium and non-equilibrium initial data.  In
\S\ref{sec:conc} we offer conclusions about this work especially
regarding its meaning for post-Newtonian and fully general
relativistic models of neutron star mergers.  

\section{Numerical Hydrodynamics Algorithms\label{sec:hydro}}

As we have mentioned in the previous section, the accuracy of the
numerical algorithms is of paramount importance if one is to obtain
accurate hydrodynamic models for mergers.  Previous work on such
models have utilized a variety of hydrodynamic schemes in either a
fixed (inertial) or rotating frame of reference.  The latter has been
claimed to be more accurate by virtue of is obviation of the
difficulties of advection, but no systematic comparison of the two has
yet been published.  In this section we describe two numerical
hydrodynamics schemes that we have employed to produce such a study.
We have not attempted an exhaustive study in which we compare the
qualitative results of each hydrodynamic scheme that has been employed
to date.  Such studies have been conducted for most of these schemes
on a number of problems involving shocks.  However, the performance of
the hydrodynamic algorithm on shocks is not the only metric by which
one needs to measure the quality of the hydrodynamic algorithm.  For
example, for simulations in which the linear momentum equation has
been solved one should examine how well angular momentum is conserved.
In the long timescale evolutions needed for multiple orbit simulations
of orbiting stars, the addition or loss of angular momentum into the
calculation could artificially enhance or delay inspiral during the
mergers.  Similar issues apply for linear momentum in simulations
where the gas angular momentum equations are solved.  In the same
vein, if the gas momentum equation is solved then one should monitor
how well the total energy is conserved.  Or if the total energy
equations is solved one should monitor how well the gas energy
equation is solved.

The 3-D numerical hydrodynamics scheme we describe in the section is
similar to the ZEUS scheme of Stone \& Norman (1982).  However, we
have made some fundamental changes to the order of operations in order
to improve the numerical accuracy of the scheme on self--gravitating
problems.

\subsection{Euler Equations}

The flow of matter in the neutron stars can be taken to be inviscid.
Under these circumstances the Newtonian description of the matter
evolution is described by the continuity equation together with the
Euler equations (Mihalas \& Mihalas 1984, Bowers \& Wilson 1991)
\nocite{mm84,bw91} of compressible inviscid
hydrodynamics.  In an inertial frame of reference the equations are
\begin{equation}
\frac{\partial \rho}{\partial t} + \nabla \cdot ( \rho {\mathbf v} ) = 0
\label{eq:cont}
\end{equation}
\begin{equation}
\frac{\partial E}{\partial t} + 
\nabla \cdot ( E {\mathbf v} ) = - P \nabla \cdot {\mathbf v}
\label{eq:gase}
\end{equation}
\begin{equation}
\frac{\partial (\rho v_i)}{\partial t} + 
\nabla \cdot ( \rho v_i {\mathbf v} ) =
- \left( \nabla P \right)_i
- \rho \left(\nabla \Phi \right)_i,
\end{equation}
where the dependent variables are the mass density $\rho$, the
internal energy density $E$, the fluid velocity $v_i$, the fluid
pressure $P$, and the Newtonian gravitational potential $\Phi$.  The
gravitational potential is described by the the Poisson equation
\begin{equation}
\nabla ^2 \Phi = 4 \pi G \rho
\label{eq:poisson}
\end{equation}
in conjunction with  boundary conditions that must be
specified.

In a frame rotating with angular frequency {\boldmath ${\omega}$} about the
center of the fixed (inertial) grid the gas momentum equation is
modified by the addition of the Coriolis and centrifugal forces
(Chandrasekhar 1969) \nocite{chandra69}
\begin{equation}
\frac{\partial (\rho v_i)}{\partial t} + 
\nabla \cdot ( \rho v_i {\mathbf v} ) =
- \left( \nabla P \right)_i
- \rho \left(\nabla \Phi \right)_i
- 2 \rho \left( \bldomega
\times {\bf v}_r \right)_i
- \rho \left( \bldomega
\times \left( \bldomega
\times {\bf r} \right) 
  \right)_i
\label{eq:rgasm}
\end{equation}
where ${\bf v}_r$ is the velocity of the rotating frame relative to
the lab frame.
The notation $({\bf A})_i$ indicates the $ith$ component of
the vector ${\bf A}$.
In the limit of $\omega \longrightarrow 0$ we recover the inertial
frame momentum equation.  
The continuity, gas energy, and Poisson
equations are unchanged from the inertial frame case.

For the remainder of this paper we will consider the angular frequency
vector $\bldomega$ to be co-aligned with the z axis so that
\begin{equation}
\bldomega = \omega \hat{{\bf e}}_z.
\end{equation}
We assume that the z axis passes through the center of the grid at
coordinates $(x_c,y_c)$.
Under this condition the Coriolis and centrifugal force terms in
Cartesian coordinates become:
\begin{equation}
\left(\frac{\partial \rho {\bf v}} {\partial t}\right)_{\mbox{cc}} = 
2\omega\rho \left( v_y \hat{\bf e}_x - v_x \hat{\bf e}_y \right)
+ \omega^2 \rho
\left(  (x-x_c) \hat{\bf e}_x + (y-y_c) \hat{\bf e}_y \right)
\label{eq:acc_cc},
\end{equation}
where $x_c$ and $y_c$ are the coordinates of the z-axis.

The set of hydrodynamic equations must be closed by specifying an
equation of state expressing pressure as function of local
thermodynamic quantities. A standard choice for building the initial
neutron star models is the polytropic equation of state, which has the
form
\begin{equation}
P = \left( \gamma -1 \right) E
\end{equation}
where $\gamma$ is the polytropic exponent, which is related to the
polytropic index, $n$, by the relationship
\begin{equation}
\gamma = 1 + \frac{1}{n}.
\end{equation}
This particular type of EOS is advantageous in that the gas energy
equation becomes linear in $E$ rendering the solution trivial.  In
isentropic situations this EOS allows the pressure to be written
purely as a function of density in the form
\begin{equation} \label{eq:polyt}
P  =  K\rho^\gamma,
\end{equation}
where $P$ is the pressure, and $K$ is the polytropic constant.  This
form of the EOS is used for the construction of the initial models.

\subsection{Numerical Solution\label{subsec:numerical}}

Much of the numerical scheme we employ for the solution of the Euler
equations is derived from the ZEUS-2D hydrodynamics scheme invented by
Stone \& Norman (1992)\nocite{sn92a} (hereafter SN).  In particular
the finite-differencing stencils are identical to those of SN, with
the exception of the Coriolis and centrifugal forces, which are not
included in the ZEUS-2D scheme.  However, the method we employ differs
in one significant way: the order of solution of the various terms in
the Euler equations differs from that of the ZEUS-2D algorithm.  As we
will show in a subsequent section of this paper, the order of the
solution of these equations is of fundamental importance to the
accuracy of the algorithm in the case of self-gravitating
hydrodynamics.  A final simplification for our algorithm, which we
shall henceforth refer to as the V3D algorithm, employs only Cartesian
coordinates.  This simplification significantly increases the
computational speed of the V3D code.

The finite-differencing algorithm we employ relies on a staggered grid
in which the intensive variables $E$, and $\rho$ are defined at cell
centers, while the vector variables such as the velocity components
$v_i$ are considered to be defined at their respective cell edges.
The gravitational potential $\Phi$ is also defined at the cell center.
The centering of the variables on the grid is depicted in a 2-D plane
of the 3-D grid in Figure \ref{fig:grid}.
\begin{figure}[tbh]

\begin{center}
\end{center}
\caption{A diagram of a 2-D slice of the staggered mesh 
illustrating the locations at which the variables are defined. The
plane shown is located at a ``z'' coordinate $z_\kpoh$. }
\label{fig:grid}
\end{figure}
In our finite-difference notation we employ superscripts to denote the
time at which the variables are defined.  The timestep is taken to be
$\Delta t$ and the numerical algorithm advances the solution of the
PDEs from $n$th time, $t^n$, to the new $(n+1)$th time, $t^\npo = t^n +
\Delta t$.

The explicit finite-difference algorithm employed by SN
for the solution of the Euler equations decomposes the time
integration into two steps by employing operator splitting among the
various terms of the equations.  In one step the density, internal
energy density, and velocities are updated by integrating the 
advective terms.  In the nomenclature of SN we refer
to this as the {\it transport} step.  The remaining terms, i.e.
the terms on the right hand sides of equations (\ref{eq:cont}),
(\ref{eq:gase}), and (\ref{eq:rgasm}), are integrated forward in time.
Following SN we refer to this step as the {\it source} step.

An additional consideration involves the solution of the Poisson
equation and how it relates to the solution of the Euler equations.
In the ZEUS-2D algorithm the order of solution of the Euler equations
is described in the flow chart of Figure \ref{fig:flow}a.  In contrast
our algorithm, V3D,
is described in Figure \ref{fig:flow}b.
\begin{figure}[h]
\begin{center}
\end{center}
\caption{The order of operations to update the hydrodynamic
and gravitational potential variables from time 
$t^n$ to time $t^{n+1}$.} 
\label{fig:flow}
\end{figure}

In the transport step the following equations are solved:
\begin{equation}
\frac{\partial \rho}{\partial t} = - \nabla \cdot ( \rho {\mathbf v} )
\label{eq:cont_t}
\end{equation}
\begin{equation}
\frac{\partial E}{\partial t} = -
\nabla \cdot ( E {\mathbf v} )
\label{eq:gase_t}
\end{equation}
\begin{equation}
\frac{\partial (\rho v_i)}{\partial t} = -
\nabla \cdot ( \rho v_i {\mathbf v} ).
\label{eq:rgasm_t}
\end{equation}
The advection integration is carried out using Norman's consistent
advection scheme (Norman 1980)\nocite{norman80}, which ties the
advected internal energy density and advected velocity to the mass
flux as described in SN.  This concept of tying the energy and
momentum fluxes to the mass flux has been shown to posses superior
angular momentum conservation properties (Norman 1980).  The actual
flux limiter employed is the van Leer monotonic flux limiter (van Leer
1977)
\nocite{vanl77} that is spatially second order.

In the source step the following equations are updated:
\begin{equation}
\frac{\partial E}{\partial t}  = - \left( P+Q \right) \nabla \cdot {\mathbf v}
\label{eq:gase_s}
\end{equation}
\begin{equation}
\frac{\partial (\rho v_i)}{\partial t}  =
- \left( \nabla P \right)_i - \left( \nabla Q \right)_i
- \rho \left(\nabla \Phi \right)_i
- 2 \rho \left( \bldomega 
\times {\bf v}_r \right)_i
- \rho \left( \bldomega 
\times \left( \bldomega 
\times {\bf r} \right) 
  \right)_i
\label{eq:rgasm_s}.
\end{equation}
The scalar viscous stress $Q$ is added to the equations in order to
allow for viscous dissipation by shocks in the fluid.  We employ the
standard von Neumann-Richtmyer prescription for the viscous stress, as
described in SN, with a length parameter of $\ell = 2$.  We monitor
the viscous dissipation arising from this stress and have found that
the total viscous energy generation is negligible for two merging
polytropes.  The coupling to gravity enters through the gradient of
the Newtonian gravitational potential in equation (\ref{eq:rgasm_s}).

We wish to note that the continuity equation is not updated during the
source step as it possesses no source term.  The lack of a source
term for the continuity equation means that the density at a new time
$t^\npo$ is known after the transport step is complete.  As we will
discuss in a subsequent section of this paper, this point is crucial
to our preferred method of solution for these equations.

The explicit finite-differencing of the Euler equations is briefly
discussed in appendix A. For the remainder of this section we
concentrate on the order of solution of the transport and source
steps.  In the method of SN the source terms (equations
\ref{eq:gase_s} and \ref{eq:rgasm_s}) are integrated forward in time
to arrive an intermediate solution for the new internal energy
density $E$ and the velocity ${\bf v}$.  The intermediate energy
density and velocity are used as initial values for the transport
equations (\ref{eq:cont_t})-(\ref{eq:rgasm_t}), which are then
integrated forward in time to find the values of the density, energy
density, and velocity at $t^\npo$.  SN have shown by means of
convergence testing that this algorithm is spatially second order
accurate.

There is no compelling reason to suggest that the order of source and
transport updates as presented by SN is preferred.  One can easily
reverse the order of updates so that the results of the transport step
are utilized in the source update.  We henceforth will refer to this
order of updates as the V3D algorithm while the opposite order will be
referred to as the ZEUS algorithm.  By comparing the two algorithms on
a number of standard hydrodynamic test problems we have numerically
verified that the reversal of these operations has no significant
effect on the overall quality of solutions when self-gravity is not
present.  However, as we will show in a later section the V3D
algorithm offers significant advantages when modeling orbiting binary
stars.

An example of the comparable performance of the ZEUS and V3D
algorithms on a standard test problem is shown in Figure \ref{fig:sod}
where the performance of the algorithm on a Sod-like (Sod 1978) 
\nocite{sod78}
shock tube is shown.
\begin{figure}[h]
\begin{center}
\end{center}
\caption{ The numerical solution of a shock tube problem for the ZEUS
(cross symbols) and V3D (circles) algorithms.  The solid line
 illustrates the exact Riemann solution to this problem.}
\label{fig:sod}
\end{figure}
The shock tube problem pictured employs a $\gamma=5/3$ polytropic
equation of state.  The grid is set up with 100 spatial zones over the
range of $-2 < x < 2$ cm with the initial contact interface at $t=0$
located at $x=0$.  This initial configuration is that of a Riemann
problem, which results in a shock and a contact discontinuity
propagating to the right and a rarefaction propagating to the left.
Because the exact solution is known (Chorin \& Marsden
1993)\nocite{chorin93} we can easily evaluate the numerical results
from both algorithms.  Overall the character of the numerical solution
is comparable between the two cases.  The values of the variables in
both the contact discontinuity and the shock are slightly different as
would be expected from different algorithms, but both methods resolve
the shock and the contact discontinuity with the same number of zones.
The rarefaction is represented nearly identically by both methods.
The figure compares the two orders of update.  One can visually see
that little difference exists between the two solutions.  We have
verified this on a number of other non-self-gravitating test problems.
In contrast, for the case of self-gravitating hydrodynamics, we do
find that the V3D algorithm is preferred as we will discuss in section
\ref{sec:self}.

\subsection{Numerical Solution of the Poisson Equation}

In order to describe self-gravitating phenomena, the gravitational
field of the matter distribution must be found by solving the Poisson
problem described by equation (\ref{eq:poisson}).  The Poisson
equation can be readily solved by a variety of techniques well suited
to elliptic equations.  In the simulations described in this paper we
have employed both W-cycle multigrid and Fast-Fourier-Transform (FFT)
methods (Press et al. 1992)\nocite{press92}.  Both methods have been
extensively tested on matter configurations where the solution is
known.  Because the Poisson equation is linear one can easily generate
test problems with known answers, but which also posses complex field
geometries.  For example, by placing $\gamma=2$ polytropes at random
points within the computational domain we can create a complex
gravitational field configuration.  Since the gravitational potential
for a $\gamma=2$ polytrope is analytically known, the potential for the
entire configuration at any point in the computational domain is
readily found as a superposition of individual polytropic solutions.
Using this method we have found that both methods give the correct
answers to approximately $\sim 10^{-5}$ for the grid resolutions
employed in this work.

The numerical solution of equation (\ref{eq:poisson}) requires the
specification of boundary values along the edge of the computational
domain.  For the problem of merging neutron stars these are {\it a
priori} unknown.  The problem of determining the appropriate boundary
conditions has been approached differently by a number of different
groups.  Ruffert et al. \nocite{rjs96} have employed zero-padding
boundary conditions in conjunction with their FFT solution method. The
zero padding method has been shown by James (1977)\nocite{james77} to
be algebraically equivalent to a direct summation by convolution of
image charges (defined on the edge of the grid) over the Green
function for the Poisson equation.  Oohara and Nakamura
(1990)\nocite{oohara90}, Rasio and Shapiro (1992,1994,1995), and New
and Tohline (1997)\nocite{nt97} have not specified how boundary
conditions on the potential were obtained for their hydrodynamic
solutions.  A number of groups (Davies et al. 1994; Zhuge et al. 1994;
Zhuge et al. 1996)
\nocite{dbpt94,zcm94,zcm96} have carried out smoothed particle
hydrodynamics (SPH) simulations in which the field was computed by a
tree-code summation obviating the need for the specification of
boundary values.

The accurate specification of self-consistent boundary conditions for
the Poisson equation is challenging.  The expansion of the potential in
terms of multi-poles may require the expansion to be carried out to
very high order if one is to obtain accurate values for the potential.
This is especially true given that the initial configuration for a
neutron star merger simulation consists of two widely separated fluid
bodies, which has a very large quadrupole moment.  The zero-padding
method necessitates a fairly large memory cost.  This memory cost can
largely be eliminated by use of the James algorithm for the solution
of the Poisson problem for isolated systems (James 1977)\nocite{james77}.  
In fact we employ the James algorithm to obtain
equilibrium initial data described in the next section.  However, we
have found that the James algorithm does not scale well to large
numbers of processors on shared-memory parallel computers.

In order to obtain an accurate algorithm for the boundary conditions
on the potential we have turned to direct integration over the Green
function for Poisson's equation, i.e.
\begin{equation}
\Phi ({\bf x}) = - \int 
\frac{G \rho({\bf x}^\prime)}
{\left|{\bf x}-{\bf x}^\prime\right|} d^3x^\prime.
\label{eq:green1}
\end{equation}
Because of the large number of grid zones present in the problem, it is
computationally intractable to compute this sum directly.  If our
computational domain is discretized into $N$ points in each of the
three spatial dimensions then the summation is over $N^3$ zones in
order to evaluate the potential at each of the $6N^2$ points on the
edge of the domain.  This implies the total algorithm is order $N^5$.
Since we typically employ $N=128$ for our simulations, this renders the
direct summation over the entire grid computationally intractable for
use in a hydrodynamic simulation.  This dilemma is further exacerbated
by the need to calculate an inverse square root in order to evaluate
the distance $r$ from the boundary point to each zone.  However, for
the problems we are considering the mass is concentrated in only a
relatively small region of the computational domain.  If we restrict
the summation to only those zones in which a significant amount of
mass is present, the summation becomes more tractable.  Accordingly, we
have adopted the following algorithm for obtaining the boundary
values.

We evaluate the amount of mass in eight-zone cubic ($2\times 2\times
2$) ``blocks'' of the grid.  If the mass of the zone is greater than a
threshold value $M_{\mbox{th}}$ the total mass of the block and the
blocks center of mass coordinates are stored in a list.  This
operation is order $N^3$ but it is only required once per timestep.
Once the entire mesh has been scanned we obtain a complete list of all
the block with a significant mass.  The size of the list is dependent
on the mass threshold employed.  If the mass threshold is chosen too
low, the list will become very large and the summation computationally
intractable.  If the cutoff is chosen too large, the list will not
include most of the mass in the domain.  We have experimentally found
that a value of $M_{\mbox{th}} = 10^{-5} M_\odot$ produces a list
which fully represents the mass in the domain.  Once the list of
significant mass blocks has been produced the boundary values of the
potential can be calculated by direct summation
\begin{equation}
\phi_{ij} = \sum^{k=L}_{k=1} \frac{M_k}{r_{ijk}},
\label{eq:bcon}
\end{equation}
where $M_k$ is the mass in the $k$th block in the list, $r_{ijk}$ is the
distance from the $ij$th point on the edge of the grid to the center
of mass coordinates of the $k$th block, and $L$ is the length of the
list.  This operation is order $N^2L$.  However, $L \ll N^3$, which
renders the summation tractable.  Furthermore, this algorithm is
readily parallelizable on a shared memory parallel computer thus
allowing for a rapid solution.  We typically find that calculation of
the boundary conditions never exceeds 20\% of the overall
computational effort.  Finally, we wish to emphasize that this
algorithm will not work efficiently in cases where the mass is more
evenly distributed over the entire mesh.

The accuracy of this algorithm has been tested by two methods.  First,
the algorithm was applied to the test problems we mentioned earlier in
this section where the analytic answer was known.  Secondly, the
summation algorithm was also tested in more general situations by
comparing the boundary values obtained by this method to those
obtained by brute force direct summation.  In all cases the boundary
values agreed to better than $10^{-4}$; in most cases the agreement
was better than $10^{-5}$.  Additionally, we track the total mass in
the list so that it can be compared to the total mass on the mesh.
Any significant difference between the two masses will indicate a
problem with the summation.

\section{Equilibrium Initial Data\label{sec:eqd}}
\subsection{Numerical Methods for Obtaining 
Equilibrium Data\label{sec:eqd_num}}

In order to accurately model two neutron stars in close orbits with
one another, it is important to employ initial conditions that
precisely reflect the true configurations of the two fluid bodies.
In general, for close binary systems, these equilibrium configurations
will not consist of two spherical stars. Instead the configuration
will contain tidally distorted fluid bodies that are only
approximately spherical.  In numerical simulations of binary star
systems the stability of the orbits can be quite sensitive to the
details of the initial configuration.  In a subsequent section we
compare dynamical models which have employed equilibrium initial
conditions with models that have utilized spherical stars.

The construction of initial data for these systems is a non-trivial
task.  In practice each neutron star in a binary system will be
non-synchronously rotating around it's own axis.  Several calculations
(Bildsten \& Cutler 1992; Kochanek 1992)\nocite{bildsten92,kochanek92} 
have shown that viscous dissipation at
the causal limit is insufficient to tidally lock binary neutron star
systems during their lifetime.  Accordingly, the most realistic
configurations that one could model would be non-tidally locked.
However, there is tremendous difficulty of obtaining equilibrium
initial data for such cases.  Finding initial conditions that
correspond to the non-synchronous case would require the solution of
the compressible Darwin--Riemann problem that is well outside the
scope of this paper.  Since the target of this paper is a study of the
numerical methods and initial conditions needed for precise
simulations of binary neutron star mergers, we restrict ourselves to
the tidally locked case.  Realistic binary systems will also contain
unequal mass components.  However, in this paper we consider only
equal mass systems.  Our numerical algorithm for obtaining
equilibrium initial data is easily extended to the non-equal mass case
which will be considered in a future paper.

A number of other research groups (Oohara \& Nakamura 1990; 
New \& Tohline 1997)\nocite{oohara90,nt97} have developed
methods to obtain equilibrium data for the case of synchronous binary
neutron star systems.  In both cases the equilibrium models must
simultaneously satisfy both the Bernoulli and Poisson equations.  In
the case of Oohara and Nakamura they have employed a method that in
similar to ours in that it iteratively solves the Bernoulli and
Poisson equations on a Cartesian grid.  However, we have found some
problems with this method for obtaining the initial conditions that we
seek.  New \& Tohline have found initial conditions using the
self-consistent field technique of Hachisu (1986a,1986b),
\nocite{hachisu86a,hachisu86b} 
which iteratively solves the Bernoulli equation and the integral form
of the Poisson equation on a spherical polar grid.  While this latter
method avoids the problems we have found with the Oohara method, it's
use for our case would involve remapping of the data from the polar
grid to the Cartesian grid we employ for our dynamical simulations.
This remapping would introduce small errors that would render the
initial conditions on the Cartesian grid slightly out of equilibrium.
In turn, the deviation from equilibrium can cause spurious
hydrodynamic motions away from the initial data once the evolution
begins.  In order to avoid this we have combined techniques from both
the Oohara et al. and New \& Tohline methods.  Our objective is to
develop equilibrium data on the same grid that the hydrodynamic
simulation will employ.

We first need to identify the equations that describe the equilibrium
configuration.  These equations result from taking the hydrostatic
limit of the Euler equations of hydrodynamics together with the
Poisson equation.   
In the hydrostatic limit the gas momentum equation
(\ref{eq:rgasm}) collapses to 
\begin{equation}
\nabla \cdot ( \rho v_i {\mathbf v} )
+ \left( \nabla P \right)_i
+ \rho \left(\nabla \Phi \right)_i
+ 2 \rho \left( {\bldomega} \times {\bf v}_r \right)_i
+ \rho \left( {\bldomega} \times \left( {\bldomega} \times {\bf r} \right) 
  \right)_i = 0.
\label{eq:rhyd1}
\end{equation}
If we make the assumption that the equation of state is of the
isentropic form given by equation (\ref{eq:polyt}), 
then equation (\ref{eq:rhyd1})
can be integrated by parts to find the Bernoulli equation
\begin{equation}
\frac{\gamma K\rho^{\gamma-1}}{\gamma-1} + \Phi + \frac{\omega^2}{2}
\left[(x-x_c)^2 + (y-y_c)^2\right] = C
\label{eq:bern1}
\end{equation}
where $C$ is a constant and where we have assumed that the rotation is
about the z-axis.  This equation must be satisfied in the interior of
the fluid bodies.  Simultaneously the equilibrium data must also satisfy
the Poisson equation
\begin{equation}
\nabla ^2 \Phi = 4 \pi G \rho.
\label{eq:poisson2}
\end{equation}
However, two fundamental difficulties occur in the solution of these
two equations on Cartesian grids containing self-gravitating fluid
bodies.  First, we do not {\it a priori} know where the boundaries of
the fluid bodies lie, and hence we do not know where the Bernoulli
condition should apply.  Second, we do not {\it a priori} know the
boundary conditions on $\Phi$ that must apply to the Poisson equation.
The boundary conditions on $\Phi$ must be determined in a
self--consistent fashion using the Green function corresponding to the
Poisson equation.  This latter problem is the more difficult of the
two problems to solve.  While Oohara et al. have employed a direct
solution of the Poisson equation for equilibrium data, they have made
no mention of what boundary conditions they have employed on equation
(\ref{eq:poisson2}).  We have found that the configuration resulting
from the iterative solution of equations (\ref{eq:bern1}) and
(\ref{eq:poisson2}) is quite sensitive to the use of
non-self-consistent boundary conditions and we strongly recommend
against employing such boundary conditions.

In order to minimize the problems with deciding where the boundaries
of the fluid bodies are, we have adopted a technique from the SCF
technique of Hachisu et al. (1990)\nocite{hachisu90}.  We consider
equilibrium binary systems in two different topological configurations
as depicted by Figure \ref{fig:binary}.  In the first case we consider
non-contact binary systems.  In the second case we consider contact
binary systems. In the first case during our iterative solution of the
combined Bernoulli and Poisson equations we specify the extremal inner
and outer points of the star as depicted in Figure \ref{fig:binary}A.
We define the orientation of our grid so that the x-axis passes
through the centers of mass of the two stars.  The z-axis passes
through the barycenter of the system thus defining the origin of the
grid.  By specifying the inner and outer points we seek equilibrium
solutions with a certain aspect ratio.  By adjusting the locations of
the extremal points we can find equilibrium configurations with
varying separations between the components and or their centers of
mass.  In the case of contact binaries we specify the extremal outer
point of the contact system and the extremal outer point of the neck
connecting the high density portions of the two fluid bodies. As with
the detached case, by varying these two points we can find a sequence
of equilibrium configurations with varying separations between their
centers of mass.  In the contact binary case we define the center of
mass of each star by considering only the mass contained within each
half of the computational domain as defined by a plane perpendicular
to the line connecting the two highest density zones of the grid.
\begin{figure}[h]
\begin{center}
\caption{Separated and contact binary topologies}
\label{fig:binary}
\end{center}
\end{figure}

Once we have identified the configuration of the system we can then
determine where the Bernoulli equation can be applied during each
iterative step.  Assuming, for
the moment, that we know $\omega$, $K$, and $C$, and that we posses
some iterative estimate of $\Phi$, when can then solve the Bernoulli
equation for a new estimate of the density $\rho$:
\begin{equation}
\rho = 
\left[\frac{\gamma-1}{\gamma K}
\left(C - \Phi -\frac{\omega^2}{2}
\left[(x-x_c)^2 + (y-y_c)^2\right]\right) 
\right]^{1/(\gamma-1)}.
\label{eq:bern2}
\end{equation}
Obviously, this equation only makes sense where the factor contained
within the square brackets of equation (\ref{eq:bern2}) is positive.  We
use this as a criterion to decide where to apply the Bernoulli
equation.  If the factor is positive the density is updated to the
density as determined by equation (\ref{eq:bern2}), otherwise the grid zone
is considered to be vacuum and the density is set equal to zero.

The determination of self-consistent boundary conditions for the
Poisson equation is of paramount importance.  In order to clarify what
we mean by the use of the term ``self-consistent'' we first clarify
the problem.  We assume that the computational domain, $\Omega$,
contains the self-gravitating fluid bodies which have compact support
within the interior of the domain, i.e. the fluid density vanishes on
the boundary of the domain $\partial \Omega$.  More simply stated, we
assume that the fluid bodies are contained inside the domain $\Omega$.
This type of self-gravitating system has been termed an ``isolated''
system (James 1977)\nocite{james77}.  Under these circumstances we
require that the boundary conditions on $\Phi$ satisfy equation
(\ref{eq:poisson2}).  Since the potential on the boundaries depends on
the density distribution $\rho({\bf r})$ we cannot {\it a priori}
self-consistently know the boundary values prior to solving the
problem.  This problem is not only relevant for initial data but is
also relevant for the solution of the Poisson problem during the
course of a self-gravitating hydrodynamic simulation.  Various groups
modeling equilibrium binary configurations have attempted to avoid
this problem.  The equilibrium sequence work of NT has utilized the SCF
method of Hachisu, which avoids this problem by employing a multi-pole
expansion of the potential in order to estimate the boundary
conditions.

In the case of isolated systems, James (1977) has shown that the
boundary conditions can be obtained exactly by use of FFT techniques.
We accordingly employ this method for use in our initial data
algorithm.  Furthermore, James has shown that this method is
algebraicly equivalent to the ``zero-padding'' technique employed by
Ruffert et al. (1996)\nocite{rjs96}.  The advantage of the James
algorithm over the zero-padding technique is that it requires
substantially less memory overhead which is a significant advantage in
a 3-D simulation.  We could also employ this algorithm to compute the
self-consistent potential during the course of our hydrodynamic
simulations.  However, in the hydrodynamic simulations we have found
it advantageous to employ equation (\ref{eq:green1}) directly to get
the boundary conditions for $\Phi$ followed by a straightforward
Poisson solve using either multigrid or FFT techniques (Press et
al. 1992)\nocite{press92}.  We have found that this method is more
amenable to implementation on the shared memory parallel computing
architectures that we employ for our simulations.

The complete algorithm for finding the initial data is as follows:
\begin{enumerate}

\item Fix inner and outer points of stars (in the detached binary)
case and outer point and neck width (in contact binary case).  Denote
the distance from the z-axis to these points as $R_{\mbox{out}}$ 
and $R_{\mbox{in}}$.

\item Make initial guess at the density distribution throughout the
computational domain.  Also guess an initial value of $K$.

\item Using density distribution solve for potential using the James
algorithm to solve the Poisson equation with self-consistent
boundary conditions.

\item Using the Bernoulli equation evaluate $\omega$ by
\begin{equation}
\omega^2 = \frac{2(\Phi_{\mbox{out}}-\Phi_{\mbox{in}})}
{R_{\mbox{out}}^2-R_{\mbox{in}}^2}
\end{equation}

\item Evaluate the Bernoulli constant, $C$, at $R_{\mbox{out}}$
using equation (\ref{eq:bern1})

\item Update $K$ by evaluating the Bernoulli equation
at some point $x$ which lies on the line through the centers of the
stars
\begin{equation}
K = \frac{\gamma-1}{\gamma} \rho^{\gamma-1}(x)
\left(C-\Phi(x)+\omega^2 (x-x_c)^2 \right)
\end{equation}

\item Calculate new value of density for every zone on the grid using
the following algorithm:
\begin{equation}
\rho({\bf x}) = \left\{
\begin{array}{ll}
\chi({\bf x})^{1/(\gamma-1)} & \mbox{if $\chi({\bf x}) > 0$}\\
0 &  \mbox{if $\chi({\bf x}) \le 0$}
\end{array}
\right.
\end{equation}
where
\begin{equation}
\chi({\bf x}) \equiv \frac{\gamma-1}{K\gamma}
\left(C-\Phi({\bf x})+\omega^2 \left[(x-x_c)^2 + (y-y_c)^2 
\right]\right).
\end{equation}

\item If the maximum relative density change in any zone of the grid is
less than $10^{-5}$ then consider the solution converged and stop.
Otherwise go to step 3.

\end{enumerate}

One major difference between this algorithm and those utilized by
others is step 6, the update of $K$.  For a particular equation of
state, e.g.  $\gamma=2$, there may not be a solution for an
equilibrium configuration with a given inner and outer point.  One can
easily see this for the case of an isolated $\gamma=2$ polytrope where
the radius is determined by (Shapiro \& Teukolsky 1983) \nocite{st83}
\begin{equation}
R = \left[ \frac{K}{2\pi G}\right]^{1/2}.
\end{equation}
In this case only a specific value of $K$ will allow the star to
``fit'' into the specified number of grid points between
$R_{\mbox{out}}$ and $R_{\mbox{in}}$.  If the value of $K$ is not
allowed to change the iterative procedure described does not converge.
In practice the change in $K$ is small.

When constructing an equilibrium sequence with fixed values of $K$, we
adjust the grid size slightly to get the desired value of $K$.  We
have found that this usually only requires changes of a few percent in
the grid spacing $\Delta x$ in order to find an equilibrium solution
for a specified value of $K$.  The total mass of the converged
equilibrium system is determined by the initial guess of the density
distribution in step 2 of the algorithm.  By multiplying the initial
guess of the density distribution by some factor we can converge to
equilibrium systems of more or less mass.  For the case of $\gamma=2$
both the Bernoulli and Poisson equations are linear in the variables
$\Phi$ and $\rho$.  In this case the total mass of the converged
solution is affected only by the initial guess at the distribution
while the value of $K$ is determined only by the grid spacing. This
renders the procedure of producing a sequence of equilibrium solutions
for a given polytropic constant, $K$, and total mass, $M_T$,
relatively easy.  In the case where $\gamma \ne 2$ the Bernoulli
equation becomes non-linear and changes in $\Delta x$ or the initial
density guess affect both the resulting value of $K$ and $M_T$.  In
this case building an equilibrium sequence becomes much more difficult
and time consuming.  For this reason we have constructed a $\gamma=3$
equilibrium sequence only at $65 \times 65 \times 65$ resolution.  We
have constructed a few specific $\gamma=3$ equilibrium configurations
at $129 \times 129 \times 129$ resolution for use in hydrodynamic
studies of stability.

\begin{figure}[h]
\begin{center}
\end{center}
\caption{ The total energy and angular momentum for equilibrium
data obtained using the James method for $\gamma=2$ ($n=1$). The
crosses indicated models constructed on a $65^3$ grid while the
squares indicate models constructed on a $129^3$ grid.}
\label{fig:eq_seq_2}
\end{figure}
\begin{figure}[h]
\begin{center}
\end{center}
\caption{ The total energy and angular momentum for equilibrium
data obtained using the James method for $\gamma=3$ ($n=0.5$).
All models were constructed on a $65^3$ grid.}
\label{fig:eq_seq_3}
\end{figure}

\subsection{Equilibrium Sequences of Initial Data\label{subsec:eqs}}

Using the method that we have described above we have constructed
equilibrium sequences of data for $\gamma=2$ ($n=1$) and $\gamma=3$
($n=0.5$) polytropes.  
For the $\gamma=2$ case, the sequences were constructed with a total
mass of $M_T = 2.8 M_\odot$ and a value of $K=4.196\times 10^4$ erg
cm$^3$ g$^{-2}$.  An isolated spherical $\gamma=2$ polytrope with these
parameters would have a radius of approximately $R \approx 10$ km and
a central density that is roughly 10 times the nuclear saturation
density of $\rho_s = 2.5\times 10^{14}$ g/cm$^3$.  Such a configuration
resembles a realistic neutron star.  The $\gamma=2$ sequence is
shown in Figure \ref{fig:eq_seq_2} while the $\gamma=3$ case is shown in
Figure \ref{fig:eq_seq_3}.  All separations shown are the center of mass
separation $a_{\mbox{cm}}$ 
which has been normalized to the spherical radius
of a single undisturbed polytrope.  Both the total energy, 
$E_{\mbox{tot}}$, and the
angular momentum, $J$, are plotted for each configuration.

In the $\gamma=2$ sequence the models with a separation of less than
about $a_{\mbox{cm}} = 2.8$ are contact binaries where the two stars are
joined by a ``neck'' of matter passing through the barycenter of the
system.  Those systems with separations greater than $a_{\mbox{cm}} > 2.8$
are detached.  For the $\gamma=3$ case the bifurcation point is at a
separation of approximately $a_{\mbox{cm}} = 3$.  In order to obtain this
number more precisely we would have to construct models at
substantially higher resolution.  Because of the difficulty of
constructing a large number of configurations 
with a specified value of $K$ and $M_T$ 
for the non-linear $\gamma=3$ case, we have chosen not to do
so.  Our purpose was to construct initial data for hydrodynamic
simulations using the same grid that we would employ for the
simulation.

For the models shown in Figures \ref{fig:eq_seq_2} and
\ref{fig:eq_seq_3} the grid resolution was approximately $\Delta x =
1.0$ km for the $65^3$ models and $\Delta x = 0.5$ km for the $129^3$
models.  From Figure \ref{fig:eq_seq_2} it is easily observed, by
comparing the $65^3$ and $129^3$ models, that the $65^3$ models do not
have adequate spatial resolution at the wider separations.
Nevertheless both the $65^3$ and $129^3$ models show minima in both
the total energy, $E_{\mbox{tot}}$, and angular momenta, $J$, 
at approximately $a_{\mbox{cm}} \approx
2.8$.  The slight variation in the data near the bifurcation point
between detached and contact binaries is due to the finite resolution
of the grid.  The contact binaries in this case may have a neck
consisting of only one or two zones, a situation which is likely to 
cause some fluctuation in both the energy and the angular momentum
due to the discrete nature of the neck.

It is interesting to compare these results with the semi-analytic work
of Lai et al.  (1993c)\nocite{lrs93c} (hereafter LRS).  LRS constructed an
equilibrium sequence of binary, compressible, Darwin ellipsoids as an
approximation to the equilibrium configurations of two synchronously
orbiting polytropes.  By identifying a turning point in the energy
versus separation curves LRS found a secular instability for
$\gamma=2$ polytropes at a separation of $a_{\mbox{cm}} = 2.76$.
Furthermore, LRS also found that these turning-points occured
simultaneously in both the total energy, $E$, and the angular
momentum, $J$.

There have also been a number of efforts to construct such equilibrium
sequences numerically.  In addition to their semi-analytic work LRS also
found equilibrium sequences obtained using a relaxation scheme which
employed smooth particle hydrodynamics methods yielded a turning point
in the energy and angular momenta for an equilibrium sequence at a
separation of $a_{\mbox{cm}}=2.9$.  In contrast New and Tohline
(1997)\nocite{nt97}, using the SCF technique, found a turning point at
$a_{\mbox{cm}}=2.98$.  Our result, which can be readily seen from Figure
\ref{fig:eq_seq_2}, yields an approximate turning point of
$a_{\mbox{cm}}\approx 2.85$.  This result is somewhat closer to the
compressible Darwin ellipsoid value and much further from the recently
obtained value of New \& Tohline.  In agreement with both LRS and NT,
we find the turning point at a point where the equilibrium systems are
still detached.  Nevertheless, the turning point is quite close to the
point at which attached systems would form.  The occurrence of a
turning point on the detached binary branch of the curve seems to
indicate that a binary system slowly spiraling inward by some energy and
angular momentum loss mechanism will encounter an instability without
ever becoming a contact binary.  The nature of this instability and
its implication for the dynamical evolution of binary systems will be
discussed in a subsequent section.

In the $\gamma=3$ case our results indicate a turning point in the
equilibrium sequence.  In this case the value of the polytropic
constant was $K=4.961\times 10^{-11}$ erg cm$^6$ g$^{-3}$ and the
total mass was $M_T = 2.8 M_\odot$.  Note that in this case the
turning point we seem to find is at approximately $a_{\mbox{cm}}
\approx 3.05$ in comparison with the LRS semi-analytic value of
$a_{\mbox{cm}} = 2.99$.  In contrast NT find a value of $a_{\mbox{cm}}
= 3.2$.  Unfortunately, Hachisu (1986b)\nocite{hachisu86b} does not
present numerical values for the separation at this turning point so
we are unable to compare to this work.  In contrast, although Rasio \&
Shapiro (1994)\nocite{rassha94} find a find an instability for the
$\gamma=3$ case of $a_{\mbox{cm}}
\approx 2.97$ based on hydrodynamic simulations, the equilibrium
sequence they obtain on the basis of relaxation methods using their
SPH code yields a turning point at $a_{\mbox{cm}} \approx 2.7$.  
This result can be contrasted with the results of NT and our own
results which show a turning point in both energy and angular
momentum at substantially larger separations.  However, the 
semi-analytic results presented in both LRS and in Lai et al.
(1994b)\nocite{lrs94b}
show a turning point at $a_{\mbox{cm}} = 2.99$.    We will discuss the 
implications of this turning point for hydrodynamic evolution of a 
binary system in a subsequent section of this paper.

\subsection{Initial Data for Hydrodynamic Models\label{subsec:initalhydro}}

Our numerical hydrodynamics method requires the density to be
non-zero everywhere on the computational grid.  Therefore, we include a
low density ($\approx$ 1 g/cm$^3$) ``atmosphere'' as a background in
regions where stellar matter is not present.  We have varied the
density between (1 - 10$^3$g/cm$^3$) in our hydrodynamic simulations
and have found that this has no discernible effect on the dynamics of
the simulations.  Other models (Ruffert et al. 1995,
1996)\nocite{rjs95} of neutron star mergers that have been carried out
with Eulerian codes have had to employ much higher densities \
($10^9$ g/cm$^3$) for the surrounding material.

However, adding matter in regions outside the stars presents
two difficulties for hydrodynamic simulations.  First, such matter
will not in general be in hydrodynamic equilibrium if it has the 
same entropy as the matter in the stars.  Thus at the
beginning of the simulation it will immediately infall towards the
stars and form an accretion shock at the surface of the stars.  This
accretion shock, while not physically troublesome because of the 
low density of the material in the atmosphere, will have the
undesirable numerical effect of driving the timestep determined
by the Courant stability condition to a very small timestep because
of the high infall velocities.  In order to counter this effect we 
make the atmosphere hot, i.e. we set the energy per baryon in the 
atmospheric material to approximately $35-40$ MeV.  This has the
effect of preventing the atmosphere from falling down onto the
neutron star surfaces.  Furthermore we
decrease this energy slightly with distance from the stars so as to
achieve a configuration that is slightly more hydrostatically stable.

A second problem originates if the atmosphere is put in place with a
non-zero velocity with respect to the stars.  If the matter is placed
on the grid with zero velocities in the lab frame, the motion of the
stars quickly sweeps up the matter into a bow shock on the front sides
of the orbiting stars.  In a circumstance similar to the accretion
shock mentioned in the previous paragraph, the bow shock has the
numerical effect of driving the Courant timestep to zero.  In order to
avoid this problem, calculations that are carried out in the lab
frame have an atmosphere with an initial 
velocity such that the material
is rotating about the center of the grid at the same speed with which
the stars are revolving.  Near the edge of the grid the velocity of
the atmospheric material is slowly tapered to zero so as to avoid a
shock forming at the edge of the grid and to keep the velocity below
the speed of light.  In the case of rotating frame simulations the
velocity of the atmospheric material is set equal to zero in the
rotating frame.

These two steps obviate the problem of having the matter accrete
onto the stars.  We wish to note that this method requires no
additional machinations to treat the material outside the stars; the
evolution of the material is described by the numerical solution
of the Euler equations.

\section{Self-gravitating Hydrodynamics Numerical 
Methods \label{sec:self}}

One of the more difficult aspects of self-gravitating hydrodynamics is
the need to self-consistently solve both the partial differential
equations describing the dynamics of the fluid and the equation(s)
that describe the gravitational field arising from that matter.  One of
the origins of this difficulty in the Newtonian case is the different
mathematical character of the two sets of equations: the Euler
equations are hyperbolic while the Poisson equation is elliptic in
nature.  While the Euler equations can be numerically solved by
explicit techniques, the Poisson equation requires an implicit
solution.  Since the two sets of equations are coupled by the
gravitational acceleration term in the gas momentum equation, one must
take care that the numerical methods employed for these coupled
equations adequately maintain all the desirable properties of the
total system such as angular momentum conservation and total energy
conservation.  In this section we compare several methods for 
these calculations that employ various methods for treating the 
gravitational acceleration term.

We wish to emphasize that no numerical scheme that solves the linear
gas momentum equations in three dimensions will guarantee the numerical
conservation of angular momentum.  The converse is also true: if one
solves the angular momentum equations in three dimensions the solution
will not in general numerically satisfy the linear momentum equations.
This discrepancy arises from the fact that the finite-differencing of
the underlying partial differential equations reduces them to
algebraic equations that must be solved for the new values of the
density, internal energy, and velocity.  Thus the five Euler equations
are sufficient to algebraicly determine the five variables.  The
finite-differencing of the angular momentum conservation equations 
will be different from the linear momentum equations and thus 
give rise to five additional equations that must be algebraicly
satisfied by the same five variables. The problem is algebraicly
over-constrained.  Despite the fact that the linear gas momentum
and linear angular momentum equations can be easily 
shown to be equivalent, i.e. that conservation of linear momentum
guarantees the conservation of angular momentum and vice versa, there
is no such equivalence between the finite-difference analogs to
these two vector equations.  A similar statement can be made about
the gas energy equation and the total energy equation.

The issue of importance to simulations of orbiting neutron stars is
how badly does numerical conservation break down over the course of
a simulation?  That is, how badly conserved are the angular momentum and total
energy over the course of a simulation?  We will consider
these issues for several different schemes for coupling the Poisson and
Euler equations.  We will also show that a superior choice among these
schemes emerges from these comparisons.  This is vital for
quantitatively accurate models of binary neutron star mergers where
it is necessary to conserve both angular momentum and energy in order
to ensure that orbital decay is physical and not the spurious result
of numerical non-conservation.

\subsection{Coupling gravity to the hydrodynamics
\label{subsec:coup}}

 As we have previously mentioned in subsection \ref{subsec:numerical},
there are two possible orders of update of the source and transport
portions of the Euler equations.  These two possibilities are
illustrated algorithmicly in Figure \ref{fig:flow}.  The ZEUS
algorithm of Stone \& Norman employs the order of update shown on the
left of Figure \ref{fig:flow} while our V3D code employs the method
shown on the right.  As we have shown in Figure \ref{fig:sod} there
is no substantive difference between these two method in the case
where self-gravity is not relevant.  However, in the self-gravitating
case these two approaches admit different possibilities for
calculating the gravitational acceleration in the gas momentum
equation.

In the case of the ZEUS algorithm, the solution of the source step
first requires the gravitational potential in order to calculate the
gravitational acceleration in the gas momentum equation.  Hence the
need for first solving the Poisson problem as described on the left
side of Figure \ref{fig:flow}.  Since the
density at the new time ($t^\npo$) is not {\it a priori} known at the
beginning timestep (at time $t^n$), the right
hand side of the Poisson equation can only be constructed using the 
density that is known at time $t^n$. Thus the Newtonian gravitational
potential is known only at time $t^n$.  Consequently the gravitational
acceleration term which is calculated from the Newtonian potential is
not time-centered between times $t^n$ and $t^\npo$.

Our code, V3D, performs the advection step before the source step that
updates the Lagrangian terms (the terms on the right hand side of the
hydrodynamics equations).  This ordering, advection before the source
update, allows the choice of computing the right-hand side of the
Poisson equation, $4\pi G \rho$, with the density at the old time step
(time lagged), the new time step (time advanced), or the average of
the two densities (time centered).  The finite-difference expressions
for these choices are:

\begin{eqnarray*}
\left(\nabla^2 \Phi \right)_{i+{\frac{1}{2}},j+{\frac{1}{2}},
k+{\frac{1}{2}}} & =
4 \pi G \rho^n_{i+{\frac{1}{2}},j+{\frac{1}{2}},
k+{\frac{1}{2}}} & \mbox{time lagged} \\
\left(\nabla^2 \Phi \right)_{i+{\frac{1}{2}},j+{\frac{1}{2}},
k+{\frac{1}{2}}} & = 
4 \pi G \rho^{n+1}_{i+{\frac{1}{2}},j+{\frac{1}{2}},
k+{\frac{1}{2}}} & \mbox{time advanced} \\
\left(\nabla^2 \Phi \right)_{i+{\frac{1}{2}},j+{\frac{1}{2}},
k+{\frac{1}{2}}} & =  2 \pi G \left(\rho^n_{i+{\frac{1}{2}},
j+{\frac{1}{2}},
k+{\frac{1}{2}}} + \rho^{n+1}_{i+{\frac{1}{2}},j+{\frac{1}{2}},
k+{\frac{1}{2}}} \right) & \mbox{time centered} \\
\end{eqnarray*}

This choice can play a significant role in the dynamics
of a simulation.  For example, in physical situations where the 
gravitational acceleration is always increasing in time,
the use of the time-lagged centering will always underestimate
the gravitational acceleration.  Over long time scales this
consistent underestimate can lead to significant deviations from
the true physical behavior of the system.  In the case of orbiting
binary stars this could lead to non-physical evolution of the orbits.
Finally, the choice of time-centering for the gravitational acceleration
can have a significant impact on the conservation of both angular
momentum and total energy.

Both total energy and angular momentum conservation are vital in 
achieving good neutron star merger models.  A lack of conservation of
either of these two quantities could lead to unphysical inspirals and 
both qualitatively and quantitatively incorrect outcomes of the
simulations.  Therefore, it is necessary for us to examine how well
both of these quantities are maintained as various choices are made
for the gravitational acceleration coupling method.

\subsection{Comparison of Gravitational Coupling Schemes in 
Rotating and Fixed Frame Calculations
\label{subsec:rot}}

In order to ascertain how the time centering affects the dynamics of
binary orbits we have conducted a simple test.  We have placed two 
spherical $\gamma=2$, $1.4 M_\odot$ polytropes in a circular orbit
with a separation of $4R_\star$ where $R_\star \approx 9.55$ km is the
radius of the spherical polytrope.  At such wide separations the
tidal distortion of the polytrope is minimal and the spherical
approximation is valid.  This latter point will be confirmed in a
subsequent section of this paper where we compare spherical and
equilibrium initial data.

In the physical situation described in the previous paragraph, the two
stars should remain in perfectly circular orbits with constant angular
momentum.  For this reason we have carried out six simulations in which
we compare the effects of the three time centerings for both the
rotating frame and inertial frame cases.  The results of these six
simulations are shown in Figure \ref{fig:sixpanels}, which depicts the
trajectories of the centers of masses of the stars, and the barycenter
of the system, in the orbital plane.  Note that the trajectories
are terminated at the point where the stars merge or where the 
simulation was stopped (if a merger did not occur).
\begin{figure}[h]
\begin{center}
\end{center}
\caption{ The trajectories of the centers of masses of both stars
and the binary system barycenter or center of mass (labeled as CM).  
The plots in the left column
correspond to rotating frame calculations while the right column
are inertial frame models.  The top row of plots are time--advanced,
the middle row are time-centered, while the bottom row is
time-retarded. The total time of each simulation is indicated at 
the top of each panel.}
\label{fig:sixpanels}
\end{figure}
The comparison among the matrix of plots reveals that the choice of
centering has a major impact on the evolution of the orbits.  The
results of the ZEUS algorithm, which employs a time--lagged centering
for the gravitational acceleration and is carried out in the inertial
frame (depicted in the bottom right panel), show a completely spurious
inspiral of the two stars in the first orbit.  In contrast, the middle
right and the top right panels show inertial frame 
models with time--centered and
time--advanced gravitational acceleration couplings.  While the decay
of the orbit is diminished with the time--centered and time--advanced
couplings, the overall evolution of the orbits is still unstable.  In
a forthcoming paper, (Calder et al. 2000)\nocite{cflsw99} 
we shall show that this is a common
feature of hydrodynamics simulations of this problem that employ
inertial frames.  The decay of the orbits in the inertial frame case
is due to the non-conservation of angular momentum.  This is shown
directly in Figure \ref{fig:jtot_comp} where the angular momentum
evolution for inertial frame models in 
the time--centered and time--advanced cases are plotted over
the first millisecond.  We have carried out additional models for each
of these cases with a series of decreasing Courant fractions.  
We define the Courant, or CFL, fraction as the ratio of our actual timestep
to the maximal possible hydrodynamic timestep as determined by the
timestep control for the ZEUS/V3D algorithm (see SN for details).
In most simulations we employ a CFL fraction of $0.4$.
The time-lagged models that we have carried out have revealed 
an even larger decrease in the angular momentum as a function of time
than the time-centered models, which explains the rapid inspiral
seen in the bottom right panel of Figure \ref{fig:sixpanels}.

As Figure
\ref{fig:jtot_converge} shows, the lack of conservation of angular
momentum is clearly related to the size of the timestep.  A perfect
algorithm would show no evolution of the angular momentum.
Additionally, while the time--lagged case is much worse than the
time--advanced case, both show a significant change in angular momentum
over the first millisecond of evolution.  In the time-lagged
case this loss
results in the decreasing orbits seen in Figure \ref{fig:sixpanels}.  
\begin{figure}[h]
\begin{center}
\end{center}
\caption{ The evolution of the angular momentum about the z-axis 
in inertial frame models for time-centered (dashed lines) and 
time-advanced (solid lines) gravity couplings for a series of CFL
fractions, $a_{\scriptsize\mbox{CFL}}$.  For the time centered
couplings the CFL fractions were: (bottom to top)
$a_{\scriptsize\mbox{CFL}}= 0.06,0.03,0.015$, and $0.0075$. For the
time advanced couplings the CFL fractions were: (top to bottom)
$0.4,0.2,0.12,0.006,0.003$, and $0.0015$.}
\label{fig:jtot_comp}
\end{figure}
In the time-advanced case the orbit outspirals and the system
eventually acquires a small drift due to the slight interactions
with the boundaries.  This drift becomes noticeable after many
orbits.
The artificial loss of angular momentum in the
calculation is due to the inability of the finite-difference scheme
to maintain conservation of both angular momentum and linear momentum.
While this loss is mitigated through the use of time-advanced
gravitational centering it is still sufficient to cause an unphysical
inspiral of the system.
\begin{figure}[h]
\begin{center}
\end{center}
\caption{ The change in the the total angular momentum (in units of 
10$^{49}$ erg seconds after 1 millisecond as a function of the CFL
fraction $a_{\scriptsize\mbox{CFL}}$ for the time-advanced (solid
line) and time-centered (dot-dashed line) schemes.}
\label{fig:jtot_converge}
\end{figure}

The use of a rotating frame helps to minimize the effects of angular
momentum loss.  With a rotating frame it is possible to choose the
angular velocity of the frame so that the motion of the stars with
respect to the frame is minimized.  The advantages of employing a
rotating frame are clearly shown in Figure \ref{fig:sixpanels}.  In
the rotating frame the advection of the stars across the grid is
minimized and the angular momentum is conserved to a much higher
degree.  Nevertheless, the time centering of the gravitational
acceleration plays a role in determining the dynamics of the orbits.
The best combination of techniques is illustrated in the top left
panel which shows the results from a simulation using both a rotating
frame and the time-advanced coupling.  This particular scheme maintains
stable orbits for the two stars for more than seven orbits at
which point the simulation was terminated.  The simulation has shown no
significant change in the orbits of the two stars over the course of
the simulation.  An examination of the angular momentum evolution for
this simulation, in Figure \ref{fig:jtot_tarf}, 
shows that the total angular momentum is well conserved.
The lines in this figure show the angular momentum contained in matter
above various density thresholds. Note that nearly half of the angular
momentum is contained in the high-density cores of the polytropes.  Also
note that the $\rho > 2.5\times 10^{14}$ g/cm$^3$ line shows that 
that there is initially a slight re-adjustment in the angular momentum
distribution as the star relaxes on the grid.  Nevertheless the total
angular momentum is fairly well conserved over the course of the
simulation.
  
In contrast the time-lagged inertial-frame case, shown in Figure
\ref{fig:jtot_tlff}, reveals poor angular momentum conservation.
This simulation shows a steady decline in the angular momentum
at all densities.  In particular the high density core has lost 
most of the angular momentum.  The loss of angular momentum terminates
approximately when the two stars have coalesced into a single
central object.  At this point the object is fairly axisymmetric
and could almost be though of as having achieved a steady state.
Under these circumstances the time centering of the gravitational
acceleration is not as critical as it was prior to coalescence and
consequently the angular momentum is better conserved at late times.

The behavior of the total energy (Figure \ref{fig:eng_tlff}) 
in the the time-lagged inertial-frame
case shows a slight decline which is not nearly so dramatic
as the behavior of the angular momentum.  Again, the decline
ceases after coalescence.  Note that Figure \ref{fig:eng_tlff}
clearly shows the transfer of gravitational potential energy
to kinetic energy during the inspiral and coalescence.  The total
internal energy changes very little throughout the coalescence.
The time-advanced rotating frame case (Figure \ref{fig:eng_tarf})
shows a very steady behavior for all of the energies, with no
substantial change throughout the length of the simulation.

Finally, we wish to point out that virtually none of the loss of
angular momentum or energy is due to dissipation by the artificial
viscosity terms in the gas energy and gas momentum equations.  The
total dissipation due to these terms is tracked throughout the
simulation, and it is many orders of magnitude below the other
energy and angular momentum scales involved.  This includes
the case where the stars have coalesced.  We find no significant
amount of shock generated dissipation as the stars merge in any of our
models.
\begin{figure}[h]
\begin{center}
\end{center}
\caption{ The evolution of the angular momentum for the
time-advanced rotating-frame case.  The lines show the total angular
momentum contained in matter above the listed density.  The line
labeled ``grid'' indicates the total angular momentum on the entire
computational grid.}
\label{fig:jtot_tarf}
\end{figure}
\begin{figure}[h]
\begin{center}
\end{center}
\caption{ The same as \ref{fig:jtot_tarf} except for the
time-lagged inertial-frame case.}
\label{fig:jtot_tlff}
\end{figure}
\begin{figure}[h]
\begin{center}
\end{center}
\caption{ The evolution of the internal, potential, kinetic, and total
energies for the time-lagged fixed-frame case.}
\label{fig:eng_tlff}
\end{figure}
\begin{figure}[h]
\begin{center}
\end{center}
\caption{ The same as \ref{fig:eng_tlff} except for the
time-advanced rotating-frame case.}
\label{fig:eng_tarf}
\end{figure}

We have assumed that the angular velocity of the rotating frame with
respect to the inertial lab frame is a constant.  Thus in situations
where the two stars inspiral due to physical processes, the stars will
acquire a non-zero velocity with respect to the rotating frame.  In
this situation one might suspect that the angular momentum
conservation might begin to break down as the stars begin moving with
respect to the grid.  However, in the next section we shall show that
the angular momentum conservation is still well--maintained even in
the case where the stars inspiral and merge.

The significant amount of energy and angular momentum non-conservation
in the fixed frame calculations clearly establish that there are
significant problems associated with their use in modeling binary
neutron stars.  We are currently surveying other hydrodynamic methods
to see if the same difficulties are present in these other schemes.
In order to avoid the problems associated with the inertial Cartesian
frames we have chosen to employ the rotating frame, time-advanced
gravitational acceleration scheme as our preferred method for 
simulating orbiting and inspiral binary neutron stars.
Using this method we turn to the study of the stability of equilibrium
models.

\section{Dynamical Studies of Newtonian Models\label{sec:stab}}

\subsection{Stability of Equilibrium Equilibrium Models}

It has been known for some time that even in the purely Newtonian
case that tidal instabilities can drive coalescence in binary
polytropic systems.  Recent semi-analytic stability analyses have been
performed by Lai, Rasio, and Shapiro (1993a,1993c,1994a,1994c) 
\nocite{lrs93a,lrs93c,lrs94a,lrs94c}
and Lai and Shapiro (1995)\nocite{laisha95}.  These models, which
treat the binary polytropes as self-similar ellipsoidal figures of
equilibrium, have found that close polytropic binary systems may be
unstable to both dynamic and secular instabilities.  In this context
we refer to a dynamical instability as one that takes place on the
orbital timescale of the binary system while secular instabilities
involve dissipative processes that may occur on much longer
timescales.  The presence of these instabilities was confirmed
numerically by Rasio and Shapiro (1992,1994,1995)
\nocite{rassha92,rassha94,rassha95} using SPH hydrodynamics methods.
More recently, New and Tohline (1997)\nocite{nt97} have performed similar
calculations using Eulerian hydrodynamics methods and have found
results for the $\gamma=2$ ($n=1$) polytropic sequences that differ
from those of Rasio \& Shapiro.  In this section we discuss our
investigations of these equilibrium sequences using the time-advanced
rotating frame hydrodynamic scheme discussed in the previous section.

The initial data for these equilibrium sequences was constructed 
as described in section \ref{sec:eqd}.  The models that we will
discuss were all run at $129 \times 129 \times 129$ resolution
with an approximate size of $65$ km in each dimension.  The grid
used to construct the equilibrium data was the same grid that 
was used for the hydrodynamic simulation, thus obviating any
introduction of error by remapping the data onto a new grid.
Since our primary interest is in neutron star mergers we have
only carried out simulations for $\gamma=2$ and $\gamma=3$
equilibrium sequences. 

The results of our simulations for the $\gamma=2$ equilibrium sequence
summarized in Figure \ref{fig:binsep_2_eql}, which shows the time
evolution of the separation between the centers of mass between the
two stars.  We have utilized the center-of-mass of the stars to
define their separation in the same fashion as LRS.  In contrast with
NT we have found pressure maxima to be ill suited for use as a
separation diagnostic since extremely small changes in the values
of the pressure in a given zone as the stars move can cause a discrete
jump in the location of the maxima.  Because the center-of-mass is
density-weighted, the location of these points changes smoothly.
\begin{figure}[h]
\begin{center}
\end{center}
\caption{ The evolution of the center-of-mass separation, 
$a_{\mbox{cm}}$ (in kilometers), 
for $\gamma=2$  equilibrium binary systems}.
\label{fig:binsep_2_eql}
\end{figure}
Note that the binary systems with initial separations greater than
approximately $28$ km seem to be stable over may orbits while those
with initial separations less than this radius do not.  Normalized to
the value of the unperturbed polytropic radius this cutoff corresponds
to a separation of $a_{\mbox{cm}} \approx 2.8$.  This is in close
agreement with the minimal energy and angular momentum separation
which was found for this $\gamma=2$ sequence (shown in Figure
\ref{fig:eq_seq_2}).  This also corresponds to the point at which
the equilibrium sequence transitions from detached to connected binary
systems.  This result is in close agreement with the predictions of
LRS.  However, while we agree with the conclusion of Rasio \& Shapiro
(1992) that models with $a_{\mbox{cm}}=2.8$ are unstable, we do not
agree with their finding that the models inspiral on a timescale of
1-1.5 times the initial orbital period.  We find that the inspiral
occurs on timescales of $3-5$ times the orbital period.  This
timescale for evolution of these close systems seems to closely follow
those of NT.  While the simulations of NT were not carried out for a
time sufficient to show instabilities, the closest separation systems
of NT showed an outward evolution comparable to ours over the first
four orbital periods.

While we seem to agree with the numerical results of NT, we disagree
with the conclusion drawn by NT regarding stability of the $\gamma=2$
polytropic sequence.  We see all systems interior to the minimum
energy separation inspiral on the timescales of several, i.e.  3-5,
orbits.  The hydrodynamic simulations of NT have stopped at four
orbital periods.  However, most of our coalescing 
systems inspiral at precisely
that time.  There is no reason to believe that the dynamical timescale
for the inspiral must only be 1-2 orbital periods.  While the
inspirals could be a result of a secular instability triggered by
numerical inaccuracies within the code, it seems more likely that the
dynamical process may take slightly longer than what is anticipated by
NT.

An interesting feature emerges from Figure \ref{fig:binsep_2_eql}
where we note that the systems with the smallest separations
spiral out slightly towards the minimum energy point before 
undergoing tidal disruption.  Similar behavior was seen by NT, 
who unfortunately terminated their calculations before the 
point where we see the inspiral occur.  This can be seen from Figure
12 of NT, which shows the growth in the moment of inertia of their
closest system.  This evolution can be interpreted as an instability
that is driving the system towards a lower energy configuration at
separations of $a_{\mbox{cm}} \approx 2.85$.

The quality of the total energy and angular momentum conservation for
the coalescing models is paramount.  As the stars coalesce a
significant amount of matter is rapidly advected about the grid even
in the rotating frame calculations.  One might suspect that the
quality of angular momentum and energy conservation might break down
under such circumstances.  However we have found that this does not
seem to happen.  This is illustrated by the results for the
$a_{\mbox{cm}} =2.78$ model which is typical of the coalescing cases.
As Figures
\ref{fig:jz_conserve} and \ref{fig:eng_conserve} indicate, both
the angular momentum and the energy are well conserved.
\begin{figure}[h]
\begin{center}
\end{center}
\caption{ The evolution of the angular momentum in the 
$a_{\mbox{cm}}=2.78$ run.
The dashed line indicates the total angular momentum in the
domain of integration as a function of time.  The dot-dashed line
indicates the loss of angular momentum from the grid obtained by
integrating the angular momentum flux over the boundary of the domain
and over time. The solid line indicates the sum of the angular momentum
in the domain plus the lost angular momentum.}
\label{fig:jz_conserve}
\end{figure}
The coalescence begins at a time of approximately 8 msec at which time
there is a substantial transfer of angular momentum from the high
density material to lower density material.  As a result of this
angular momentum transfer and the disruption of the stars 
tidal ``arms'' are formed of material that is stripped from 
the stars.  These tidal arms contain a significant fraction
of the total angular momentum.  Some of the material in these 
arms is swept off of the grid, carrying with it angular momentum.
In Figure \ref{fig:jz_conserve} we separately track the total
angular momentum on the grid at every instant in time along with the
cumulative total of the angular momentum swept off of the grid.  The
total angular momentum is the sum of these two curves.  The components
of the angular momentum displayed in Figure  \ref{fig:jz_conserve} are
entirely composed of angular momentum about the z-axis; the x and y
components are effectively zero.  The angular momentum on the grid
undergoes a sharp decline during the merger as matter flows off the
grid.  This is also reflected in the rise of the cumulative total
angular momentum that has been advected off the grid by the matter.
Yet the total angular momentum remains quite well conserved.  We wish
to emphasize that we have not adjusted the rotation speed of the frame
as the stars have inspiraled; the grid has maintained a constant
rotation speed with respect to the laboratory inertial frame.  The
introduction of a time-varying grid rotation speed would complicate 
the hydrodynamic equations and would at best yield only relatively
small improvements in angular momentum conservation.
\begin{figure}[h]
\begin{center}
\end{center}
\caption{ The evolution of the potential, kinetic, internal, and total
energies for the $a_{\mbox{cm}}= 2.78$ run}
\label{fig:eng_conserve}
\end{figure}
The evolution of the three components of the total energy is shown in
\ref{fig:eng_conserve}.  There is a slight rise of a few percent in
the total energy over the course of the entire simulation but no
significant jump during the coalescence.  A modest transfer of
kinetic and potential energy occurs during the the merger but this
does not have a pronounced effect on the conservation of total energy.

The hydrodynamic evolution of models from the $\gamma=3$ equilibrium
sequence shows an instability at a separation of approximately
$a_{\mbox{cm}} \approx 2.85$ in good agreement with the minimum energy
and angular momentum separation of $a_{\mbox{cm}} =3.0$.  The
evolution of three models with $129 \times 129 \times 129$ resolution
is shown in Figure \ref{fig:binsep_3_eql}
\begin{figure}[h]
\begin{center}
\end{center}
\caption{ The evolution of the center-of-mass separation, $a_{\mbox{cm}}$, 
for $\gamma=3$ equilibrium binary systems}.
\label{fig:binsep_3_eql}
\end{figure}
where the binary center-of-mass separation is shown.  Because of the
difficulty of constructing high-resolution equilibrium models for the
$\gamma=3$ sequence, we have carried out only five simulations
bracketing the predicted point of instability.  Our results again
agree with the location of the instability identified in RS94 (see
RS94 Figure 3).  RS94 found the instability occured at $a_{\mbox{cm}}
=2.97$, a value within 10\% of the semi-analytic prediction of LRS of
a point of instability of $a_{\mbox{cm}}=2.7$.  In contrast, we do not
agree with the results of NT who find that binaries at larger
separations, e.g.  $a_{\mbox{cm}}=3.1$ models, are unstable to merger
(see Figure 13 of NT).  Our models with this initial separation
exhibit no sign of instability.  A puzzling fact about the NT results
for the $\gamma=3$ sequence is that even the largest separation model
with an initial value of $a_{\mbox{cm}}=3.41$ show signs of a slow
orbital decay.  Neither we, nor RS, see such behavior.

\subsection{Comparison of Models Using Equilibrium and 
Non-equilibrium Data\label{subsec:eqcomp}}

Many numerical investigations of the dynamics of binary 
neutron star coalescence have employed spherical stars as initial data.
As we discussed in section \ref{subsec:rot} the tidal distortions
for widely separated stars are small and one can often assume that
spherical stars are a good approximation to the true equilibrium
fluid bodies.  This assumption clearly breaks down as the separation
between the two stars is reduced.  The critical question is as what
point does this break down occur?

In order to clarify the realm of validity of the spherical initial data
approximation we have carried out a series of simulations using
$\gamma=2$ and $\gamma=3$ polytropes as initial data.  The initial
separations were varied in the same fashion as the the series of runs
for equilibrium data models.  The evolution of these separations for
the $\gamma=2$ case is shown in Figure \ref{fig:binsep_2_sph}.
\begin{figure}[h]
\begin{center}
\end{center}
\caption{ The evolution of $a_{\mbox{cm}}$ for $\gamma=2$ 
spherical-star binary systems}.
\label{fig:binsep_2_sph}
\end{figure}
At larger separations the systems are stable for long timescales; the
separations to do not significantly evolve.  The small oscillations
present reflect the fact that the spherical stars are not in perfect
equilibrium initially and consequently the evolving systems undergo
small epicyclic oscillations.  Similar behavior has been seen by 
RS94.  For binary separations nearer the equilibrium sequence stability
limit we can see substantial differences between the $\gamma=2$
binaries shown in Figure \ref{fig:binsep_2_sph} and their equilibrium
counterparts shown in Figure \ref{fig:binsep_2_eql}.  For systems with
separations less than 30 km, the orbital separation is diminishing.  In
the equilibrium case these systems are stable as is seen in Figure 
 \ref{fig:binsep_2_eql}. Similar behavior is seen in the $\gamma=3$
case as is shown in Figure \ref{fig:binsep_3_comp}, which compares
the equilibrium and spherical-star models.
\begin{figure}[h]
\begin{center}
\end{center}
\caption{ A comparison of the  evolution of $a_{\mbox{cm}}$ for $\gamma=3$ 
equilibrium and spherical-star binary systems}.
\label{fig:binsep_3_comp}
\end{figure}
The rate at which systems inspiral is clearly high for systems of
smaller initial separation.  In fact, the closest systems are
disrupted almost immediately.  However, the systems with initially
wider separations show no sign of instability.

The results clearly indicate that for systems with initial separations
well beyond the tidal instability limit that the spherical-star
approximation is quite acceptable.  This point is very important for 
the case of the coalescence of two rotating neutron stars where one 
would have to solve the
compressible Darwin--Riemann problem in order to obtain equilibrium 
initial data.  By starting sufficiently far beyond the tidal
instability limit one may be able to effectively employ static models
of isolated rotating neutron stars as initial data for binary
configurations.  Furthermore, the isolated star approximation, 
using post-Newtonian models for the isolated stars,
could also greatly simplify the construction of initial data for
post-Newtonian simulations as well.

\section{Conclusions\label{sec:conc}}

Self-gravitating hydrodynamic models for binary neutron star 
phenomena pose some unique challenges for numerical modelers.
Since the use of Eulerian hydrodynamics techniques is prevalent 
in Newtonian, post-Newtonian, and relativistic models of 
binary neutron star coalescence, it is vital that we have a good
understanding of the role that the numerical techniques play in 
determining the outcome of the models.  To this end, we have 
carried out a number of studies designed to compare rotating
and inertial frame Newtonian hydrodynamic models as well as to compare
several choices that could be made for the coupling of gravity
and matter.  The lessons that are learned from this efforts
will be invaluable for post-Newtonian and relativistic models
as well.  

We have been able to show that a combination of a rotating frame of
reference and a time-advanced gravitational acceleration centering
in the gas momentum equation yields adequate angular momentum
conservation for the orbiting and merging binary problems.
In contrast, we have found that the use of a inertial laboratory
frame together with a time-lagged gravitational coupling yields
incorrect results.  The inertial frame methods produces a substantial
angular momentum loss that leads to a spurious inspiral of what should 
be a stable Newtonian binary system.  This result indicates that the
use of inertial frames which involve stars advecting across the grid
should be avoided where possible.

We have found a reliable method for constructing equilibrium initial
data on the hydrodynamic grids for use in hydrodynamic simulations.
This method employs a method of solving the Poisson problem, including
the determination of consistent boundary conditions, for isolated
self-gravitating systems without having to resort to multi-pole
expansions of the mass distribution.  This iterative method is 
easily implemented and produces initial data that is consistent with 
the hydrodynamic grid.  Using this method we have constructed
equilibrium sequences that closely agree with the semi-analytic 
calculations of Lai, Rasio, and Shapiro for $\gamma=2$ and $\gamma=3$
polytropic sequences.  While we see qualitative
similarities with the results of New \& Tohline the locations of 
the minima differ somewhat from theirs and are closer to the LRS
predictions.

Using the initial data from our equilibrium sequences we have
investigated the stability of these models.  Our results are 
in very close agreement with the numerical SPH models 
of Rasio and Shapiro.  In contrast, we find that we disagree with 
the conclusions of New \& Tohline on the stability of 
the $\gamma=2$ and $\gamma=3$ equilibrium sequences.

Finally, we investigate the effects of using the isolated star
approximation for initial data.  We find that for separations 
modestly greater than the tidal instability limit that the use
of isolated polytropes for initial data has little influence 
on the subsequent evolution of the binary system.  This point
justifies the use of the isolated star approximation for the 
construction of equilibrium data for rotating, post-Newtonian,
and other complex binary neutron star systems.

\section{Acknowledgments}

We would like to express our thanks to our colleagues P. Annios,
G. Daues, J. Hayes, I. Iben, E. Seidel, P. Saylor, M. Norman,
W.-M. Suen, I. Foster, J. Lattimer, P. Leppik, M. Prakash,
P. Marronetti, G. Mathews, J. Wilson, and D. Mihalas for many helpful
conversations regarding this work.  We would also like to thank John
Shalf, David Bock, John Bialek, and Andy Hall for extensive
visualization support.  Finally we wish to thank the NASA Earth and
Space Science High Performance Computing and Communications Program
for funding for this work under NASA CAN S5-3099.  ACC would also like
to thank the Department of Energy under grant No. B341495 to the
Center for Astrophysical Thermonuclear Flashes at the University of
Chicago.  Computational resources were provided by the National Center
for Supercomputing Applications under Metacenter allocation
\#MCA97S011N.


\appendix
\section{Hydrodynamic Algorithm}

In this appendix we briefly describe the details of our hydrodynamic
algorithm.  The finite-differencing methodology is identical to that
for the ZEUS algorithm as described by Stone \& Norman (1992).  An
exception is the addition of the Coriolis and centrifugal force terms,
which do not appear in SN.  Additionally, we have employed only
Cartesian coordinates, which significantly simplifies the equations as
compared to the generalized coordinates of SN.  For intimate details
of the algorithm we refer the reader to SN.

 We wish to point out one important difference of our algorithm (V3D)
from the ZEUS algorithm: although the finite-difference stencils 
of the equations
are identical, the order of updates differs significantly as discussed in
section \ref{subsec:numerical}.   Nevertheless, we still employ a multi-step
(operator split) methodology as discussed in section \ref{sec:hydro}.
In the SN nomenclature the Euler equations are broken up into the
{\it transport} and {\it source} terms.  The transport step results in the
update of the hydrodynamic variables as due to the advective terms of
the Euler equations, while the source step results in the updates due to
the remaining terms.  We describe each of these in turn.

\subsection{Transport Step \label{subsec:transport}}

In the transport step, equations (\ref{eq:cont_t})-(\ref{eq:rgasm_t})
are updated.  These equations represent only the advective part of the
hydrodynamic evolution.  Because of the 3-dimensional nature of these
equations we employ the widely used dimensional operator splitting
technique of Strang (1968)\nocite{strang68} which decomposes the 3-D
update into a series of 1-D updates in each dimension.  For simplicity
we will describe our updates in only the x-direction.  The application
to the y- and z-directions is obvious.

In each dimension equations (\ref{eq:cont_t})-(\ref{eq:rgasm_t}) are
simple conservation laws of the form
\begin{equation}
\frac{\partial q}{\partial t} + \frac{\partial F(q)}{\partial x} = 0
\end{equation}
where $q$ is generic variable representing and advected quantity and
$F(q)$ is the flux of that quantity in the x-direction.  For the five
equations (\ref{eq:cont_t})-(\ref{eq:rgasm_t}) $q$ takes the
respective forms of $\rho$, $E$, or $\rho v_i$, while $F(q)$ takes the
forms of $\rho v_x$, $E v_x$, and $\rho v_i v_x$.  The fluxes are
calculated for the x-faces of a cell centered around the point at
which $q$ is defined.  Note that these cells will differ for the five
variables with the exception of $\rho$ and $E$ which are both defined at 
the same point.  For a given cell the update of $q$ will take the form
\begin{equation}
\tilde{q} = q^n+\Delta t (F_l(q)-F_r(q))/ \Delta x
\end{equation}
where $F_r$ and $F_l$ are the fluxes on the right and left x-faces of
the cell, $\Delta X$ is the cell width, and $q^n$ is the known value
of $q$ at timestep $t^n$.  The notation $\tilde{q}$
is used to denote that this is only a partial update of $q$ only due
to advection.  The advection scheme utilizes the consistent advection
scheme of Norman (1980)
\nocite{norman80} which ties the energy and momentum fluxes to the mass
flux, i.e. we define
\begin{equation}
F(E) = \varepsilon^\star F(\rho),
\end{equation}
where $\varepsilon$ is the internal energy per gram and
\begin{equation}
F(\rho v_i) = v_i^\star F(\rho),
\end{equation}
where
\begin{equation}
F(\rho) = (\rho v_x)^\star
\end{equation}
where the $\star$ indicates that values of variables are calculated
using the monotonic advection scheme of van Leer (1977)\nocite{vanl77}.
The implementation of
this scheme is detailed in SN and we refer the reader to that paper
for more information.

\subsection{Source Step \label{subsec:source}}

The implementation of the source step in nearly identical to that of
SN.  In this step we solve equations (\ref{eq:gase_s}) and
(\ref{eq:rgasm_s}).  Note that there are no source terms for the
continuity equation and thus the source step does not affect any
change in the density.  Thus the updates of the internal energy
density, $E$, and the momentum density components, $\rho v_i$, 
are of the form
\begin{equation}
q^{n+1} = \tilde{q} +\Delta t (\mbox{source terms}).
\end{equation}
This update makes use of the intermediate result obtained from the
previously undertaken transport step.  Since the details of the
finite-differencing for most of these source terms are given in SN we
refer the reader to that work for further information.  However, in
the gas momentum equation the the Coriolis and centrifugal force terms
are differenced as
\begin{eqnarray}
(v_x)^{n+1}_{i,j+1/2,k+1/2} & = & (\tilde{v}_x)_{i,j+1/2,k+1/2}+ 
\omega^2 \Delta t (x_i-x_{\scriptsize\mbox{c}}) +\frac{\omega
\Delta t}{2} \left[ \right. \nonumber \\ & &
\left. (v_y)^n_{i+1/2,j+1,k+1/2} + (v_y)^n_{i+1/2,j,k+1/2} \right.
\nonumber \\
& &
  + \left.(v_y)^n_{i-1/2,j+1,k+1/2} +  (v_y)^n_{i-1/2,j,k+1/2}  \right] \\
\label{eq:candcx}
\end{eqnarray}
where $x_{\scriptsize\mbox{c}}$ is the x-coordinate of the grid
center.  In equation (\ref{eq:candcx}) we have employed the difference
notation as detailed in SN.  The update for the y-velocity component
is similarly given by
\begin{eqnarray}
(v_y)^{n+1}_{i+1/2,j,k+1/2} & = & 
(\tilde{v}_y)_{i+1/2,j,k+1/2}+ \omega^2 \Delta t 
(y_j-y_{\scriptsize\mbox{c}}) 
-\frac{\omega \Delta t}{2} \left[ \right. \nonumber \\
& & \left. (v_x)^n_{i+1,j+1/2,k+1/2} +  (v_x)^n_{i+1/2,j-1/2,k+1/2} \right. 
\nonumber \\
& &
  + \left.(v_x)^n_{i,j+1/2,k+1/2} +  (v_x)^n_{i,j-1/2,k+1/2}  \right]
\label{eq:candcy}
\end{eqnarray}
where $y_{\scriptsize\mbox{c}}$ 
is the y-coordinate of the center of the grid.  Note that
since we are considering non-inertial frames that rotate around the
z-axis, there are no Coriolis or centrifugal contributions to the
z-component of the momenta.  The Coriolis and centrifugal updates
are completed in the middle of the source step. After the updates
to the momenta due to pressure and gravitational accelerations
have been completed, the velocities are calculated from the momenta.
Equations (\ref{eq:candcx}) and (\ref{eq:candcy}) are then applied to 
get the non-inertial force updates to the velocities.  Finally, the
viscous stress updates to the velocities are completed.  As the last step 
the source terms for the gas energy equation are solved in order to obtain
the new internal energy.


\end{document}